\documentclass{emulateapj}  

\usepackage{rotating}
\usepackage{color}
\usepackage[breaklinks]{hyperref}
\usepackage{breakurl}
\slugcomment{Accepted to The Astrophysical Journal Letters}
\usepackage{graphicx,amssymb,amsmath,times,subfigure}
\usepackage{natbib}
\usepackage[]{xcolor}
\usepackage{etoolbox}

\makeatletter

\def\ltsima{$\; \buildrel < \over \sim \;$}
\def\simlt{\lower.5ex\hbox{\ltsima}} 
\def\gtsima{$\; \buildrel > \over \sim \;$}
\def\simgt{\lower.5ex\hbox{\gtsima}} 
\def\arcsec{\hbox{$^{\prime\prime}$}}

\def\p0{$\pi^{\rm 0}$}

\def\r95{$r_{\rm 95}$}

\newcommand{\fermi}{\emph{Fermi}~}

\newcommand{\pg}{PG$~$1553+113}

\begin{document}

\vskip -0.8cm

\title{Multiwavelength Evidence for Quasi-periodic Modulation\\ in the Gamma-ray Blazar PG 1553+113}

\shorttitle{Quasi-periodic modulation in PG 1553+113}
\shortauthors{Ackermann et al.}

\author{ %
\small{
M.~Ackermann\altaffilmark{1},
M.~Ajello\altaffilmark{2},
A.~Albert\altaffilmark{3},
W.~B.~Atwood\altaffilmark{4},
L.~Baldini\altaffilmark{5,3},
J.~Ballet\altaffilmark{6},
G.~Barbiellini\altaffilmark{7,8},
D.~Bastieri\altaffilmark{9,10},
J.~Becerra~Gonzalez\altaffilmark{11,12},
R.~Bellazzini\altaffilmark{13}
% ...
} 
}
\altaffiltext{1}{Deutsches Elektronen Synchrotron DESY, D-15738 Zeuthen, Germany}
\altaffiltext{2}{Department of Physics and Astronomy, Clemson University, Kinard Lab of Physics, Clemson, SC 29634-0978, USA}
\altaffiltext{3}{W. W. Hansen Experimental Physics Laboratory, Kavli Institute for Particle Astrophysics and Cosmology, Department of Physics and SLAC National Accelerator Laboratory, Stanford University, Stanford, CA 94305, USA}
\altaffiltext{4}{Santa Cruz Institute for Particle Physics, Department of Physics and Department of Astronomy and Astrophysics, University of California at Santa Cruz, Santa Cruz, CA 95064, USA}
\altaffiltext{5}{Universit\`a di Pisa and Istituto Nazionale di Fisica Nucleare, Sezione di Pisa I-56127 Pisa, Italy}
\altaffiltext{6}{Laboratoire AIM, CEA-IRFU/CNRS/Universit\'e Paris Diderot, Service d'Astrophysique, CEA Saclay, F-91191 Gif sur Yvette, France}
\altaffiltext{7}{Istituto Nazionale di Fisica Nucleare, Sezione di Trieste, I-34127 Trieste, Italy}
\altaffiltext{8}{Dipartimento di Fisica, Universit\`a di Trieste, I-34127 Trieste, Italy}
\altaffiltext{9}{Istituto Nazionale di Fisica Nucleare, Sezione di Padova, I-35131 Padova, Italy}
\altaffiltext{10}{Dipartimento di Fisica e Astronomia ``G. Galilei'', Universit\`a di Padova, I-35131 Padova, Italy}
\altaffiltext{11}{NASA Goddard Space Flight Center, Greenbelt, MD 20771, USA}
\altaffiltext{12}{Department of Physics and Department of Astronomy, University of Maryland, College Park, MD 20742, USA}
\altaffiltext{13}{Istituto Nazionale di Fisica Nucleare, Sezione di Pisa, I-56127 Pisa, Italy}
% ...
%
%
%
%
%
%
%
\begin{abstract}
%
%
%
%
%
We report for the first time a $\gamma$-ray and multiwavelength nearly-periodic oscillation in an active galactic nucleus.
Using the \textit{Fermi} Large Area Telescope (LAT) we have discovered an apparent quasi-periodicity in the $\gamma$-ray flux ($E>100$ MeV) from the GeV/TeV BL Lac object PG$~$1553+113.
%
The marginal significance of the $2.18 \pm 0.08$ year-period $\gamma$-ray cycle
is strengthened by correlated oscillations observed in radio and optical fluxes, through data collected in the OVRO, Tuorla, KAIT, and CSS monitoring programs and \textit{Swift} UVOT. The optical cycle appearing in $\sim 10$ years of data has a similar period, while the 15 GHz oscillation is less regular than seen in the other bands.
Further long-term multi-wavelength monitoring of this blazar may discriminate among the possible explanations for this quasi-periodicity.
%
%
%
%
%
%
%
%
%
%
%
\end{abstract}

\keywords{gamma rays: galaxies --- gamma rays: general --- BL Lacertae objects: general  ---  BL Lacertae objects: individual (PG 1553+113) --- galaxies: jets --- accretion, accretion disks}


\section{Introduction} \label{sect:intro}

Among active galactic nuclei (AGN),  blazars are distinguished by erratic variability at all energies on a wide range of timescales.
They are generally thought to be powered by supermassive black holes (SMBHs, 10$^8$--10$^9$ M$_{\odot}$).  \pg\ \citep[1ES$~$1553+113,  $z \sim 0.49$,][]{danforth10,aliu15,sanchezhess2014}
%
%
is an optically/X-ray selected BL Lac object \citep{falomo90} emitting variable GeV/TeV $\gamma$ radiation \citep{magicpaper14,sanchezhess2014}. As typical in very-high energy (VHE) BL Lacs, the energetic non-thermal emission of \pg\ originates in a relativistic jet and has a spectral energy distribution (SED) with two humps,
overwhelming any other component from either the nucleus or the host galaxy.
%
%

The Large Area Telescope (LAT) on the {\it Fermi Gamma-ray Space Telescope} is providing continuous monitoring of the high-energy $\gamma$-ray sky. The apparent modulation noted in the $\gamma$-ray flux of \pg\ stimulated the multi-frequency
and long-term variability study described in this paper.

In \S \ref{sect:data} we describe the \fermi LAT data analysis and the sources of multiwavelength data; \S  \ref{sect:timing} details the multiple approaches used for lightcurves and cross-correlation analysis; \S  \ref{sect:discussion} outlines preliminary scenarios
to interpret these results.

\begin{figure*}[htt!!]
\hspace{-1.0cm}
\resizebox{1.15\hsize}{!}{\rotatebox[]{0}{\includegraphics{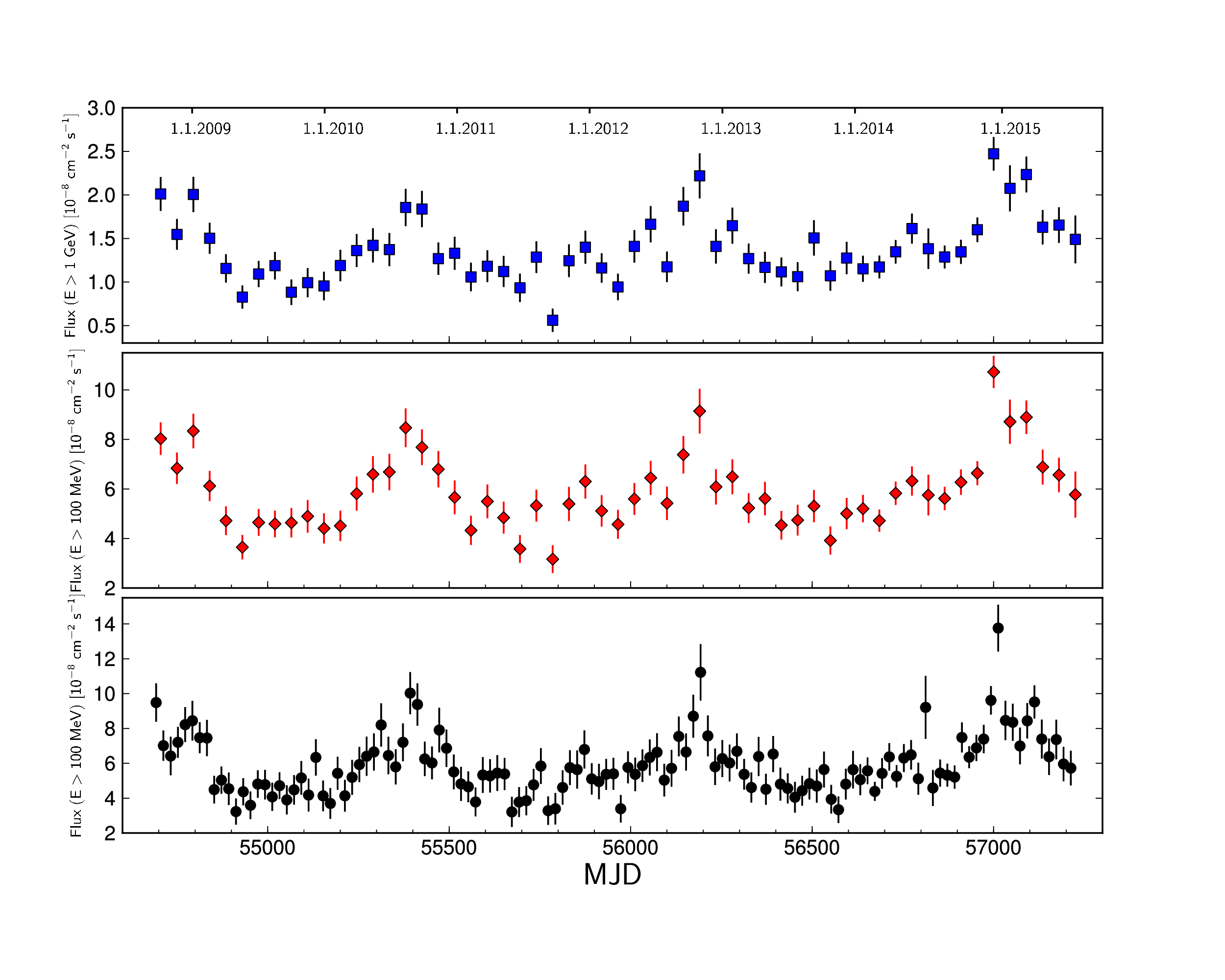}}}
\vskip -1.5cm \caption{
\fermi LAT $\gamma$-ray lightcurves of \pg\ over $\sim$ 6.9 years, from 2008 August 4 to 2015 July 19. The lightcurve above 1 GeV is shown with a constant 45-day binning (top panel); two light curves above 100 MeV are shown, with 45- and 20-day binning (middle and bottom panels)
%
%
}
\label{fig:latlc}
\vskip 0.3cm
\end{figure*}

%
\begin{figure*}[htt!!]
\vskip -1.0cm
\hspace{-1.0cm}
%
\resizebox{1.15\hsize}{!}{\rotatebox[]{0}{\includegraphics{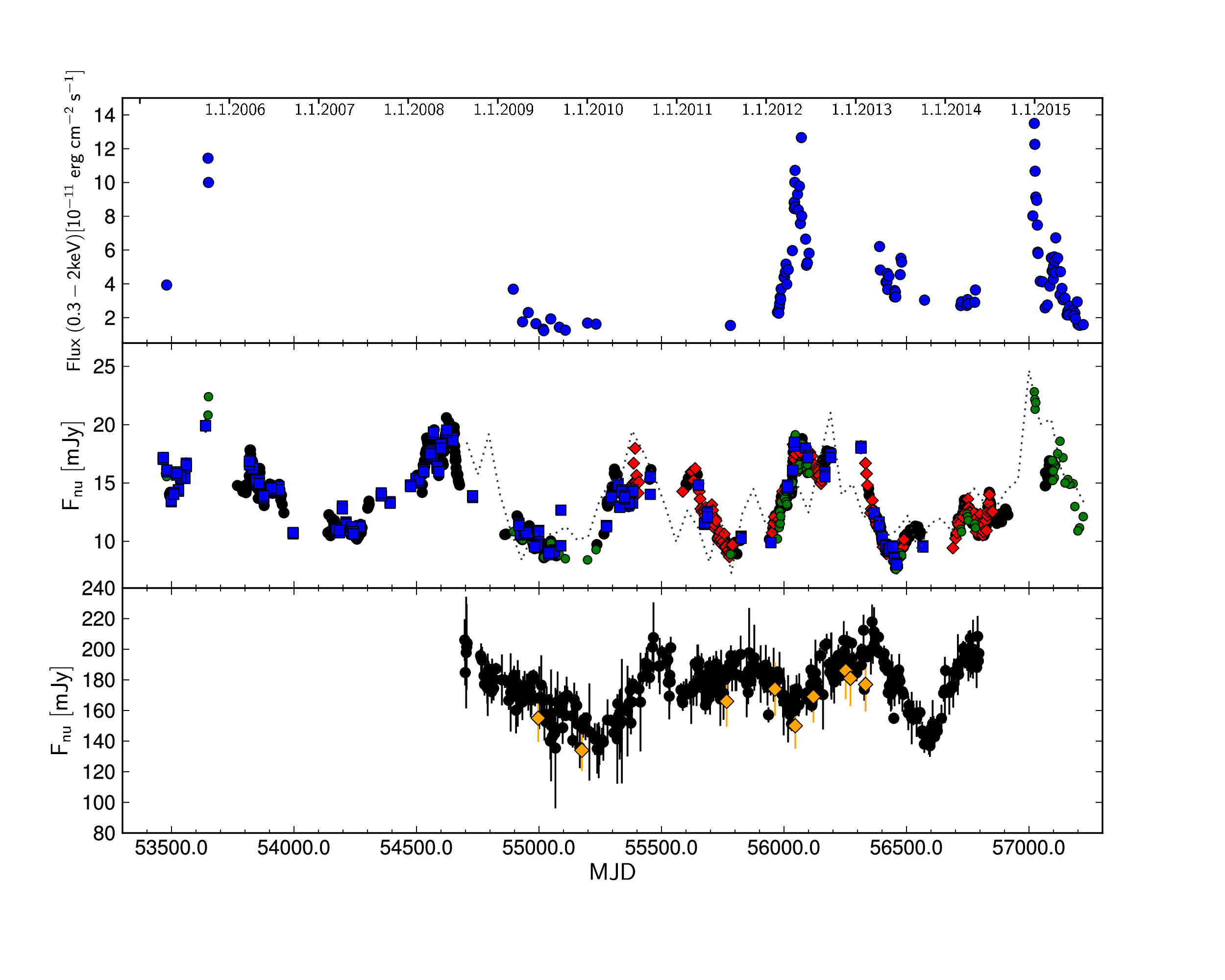}}}
\vskip -1.4cm
\caption{
\footnotesize{
Multifrequency lightcurves of \pg\ at X-ray, optical and radio bands. Top panel: \textit{Swift}-XRT integrated flux (0.3-2.0 keV).
Central panel:  optical flux density from Tuorla (R filter, black filled circle points), Catalina CSS (V filter rescaled, blue filled squared points), KAIT (V filter rescaled, red filled diamond points), \textit{Swift}-UVOT (V filter rescaled, green filled circle points). Dotted line:  LAT flux ($E > 100$ MeV) with time bins of 20 days, scaled and y-shifted for comparison.
Bottom panel: 15 GHz flux density from OVRO 40m (black filled circle points) and parsec-scale 15 GHz flux density by VLBA (MOJAVE program, yellow diamond filled points). \vskip 0.3cm
}
}
\label{fig:multifreqlc}
\end{figure*}
%
%

\section{Fermi LAT and Radio, Optical, X-ray Data} \label{sect:data}

The LAT is a pair conversion  detector with a 2.4 sr field of view,  sensitive to $\gamma$ rays from $\sim 20$ MeV to $>300$ GeV \citep{atwood09}.
%
%
The present work uses the new Pass 8 LAT database \citep{atwood13}.
The LAT operating mode allows it to cover the entire sky every two $\sim$1.6-hour spacecraft orbits, providing a regular and uniform
view of  $\gamma$-ray sources, sampling timescales from hours to years.
This work uses observations of \pg covering $\sim 6.9$ years
(2008 August 4 to 2015 July 19, Modified Julian Day, MJD, 54682.65--57222.65). 
The LAT data analysis employed the standard \texttt{ScienceTools} v10r0p5\footnote{\texttt{http://fermi.gsfc.nasa.gov/ssc/data/analysis\\/documentation/}} package, selecting events from 100 MeV$-$300 GeV 
with {\tt P8R2\_SOURCE$\_$V6} instrument response functions, in a circular Region of Interest of 10$^\circ$ radius centered on the position of \pg. It used files \texttt{gll\_iem\_v06} and \texttt{iso\_P8R2\_SOURCE\_V6\_v06} to model the Galactic and isotropic diffuse emission. Contamination due to the $\gamma$-ray-bright Earth limb is avoided by excluding events with zenith angle $>90^\circ$.
An unbinned maximum likelihood model fit technique is applied to each time bin with a power-law spectral model and photon index fixed to the 3FGL Catalog average value \citep[$1.604\pm 0.025$, ][]{acero15} for \pg. The resulting lightcurves are shown in Fig. \ref{fig:latlc}.

Optical R-band data covering an interval of $\sim 9.9$ years (2005 April 19 to 2015 March 29, MJD 53479-57110) are reported in Fig. \ref{fig:multifreqlc}. Most unpublished observations were performed as part of the
Tuorla blazar monitoring program \citep{takalo08}\footnote{\texttt{users.utu.fi/kani/1m}}.
%
%
%
%
The data are reduced using a semi-automatic pipeline (Nilsson et al. in prep.). 
%
%
Public data from the Katzman Automatic Imaging Telescope (KAIT) and the Catalina Sky Survey (CSS) programs are also added. V-band magnitudes are scaled
to the R-band values.
%
%
%

As part of an ongoing blazar monitoring program supporting \fermi\ \citep{ovromonitorprogram}, the Owens Valley
Radio Observatory (OVRO) 40-m radio telescope has been observing \pg\ continually (about every  1 to 23 days) since 2008 August. Figure~\ref{fig:multifreqlc} reports published 15~GHz lightcurves for the period from 2008 August 19 to 2014 May 18 (MJD 54697-56795).
OVRO instrumentation, data calibration and reduction are described in \citet{ovromonitorprogram}.
%
%
%

{\it Swift} observed \pg\ 110 times between 2005 April 20 and 2015 July 18 (unabsorbed 0.3--2 keV flux lightcurve in Figure \ref{fig:multifreqlc}).
X-Ray Telescope  (XRT) data were first calibrated and cleaned
({\tt xrtpipeline}, XRTDAS v.3.0.0) and energy spectra extracted from a
region of 20 pixel ($\sim$47 $\arcsec$) radius, with a
nearby 20 pixel radius region for background.
Individual XRT spectra are well fitted with a log-parabolic model,
with
column density fixed to the Galactic value 
of $3.6 \times 10^{20}$ cm$^{-2}$ \citep{kalberla05}.
%
%
Aperture photometry (5 $\arcsec$ radius) for the UVOT V-band filter was performed.
%
%
%
%

\begin{figure*}[hhhhtt!!]
\vskip -0.2cm
\hspace{-0.9cm}
\begin{minipage}[c]{0.5\linewidth}
\resizebox{1.0\hsize}{!}{\rotatebox[]{0}{\includegraphics{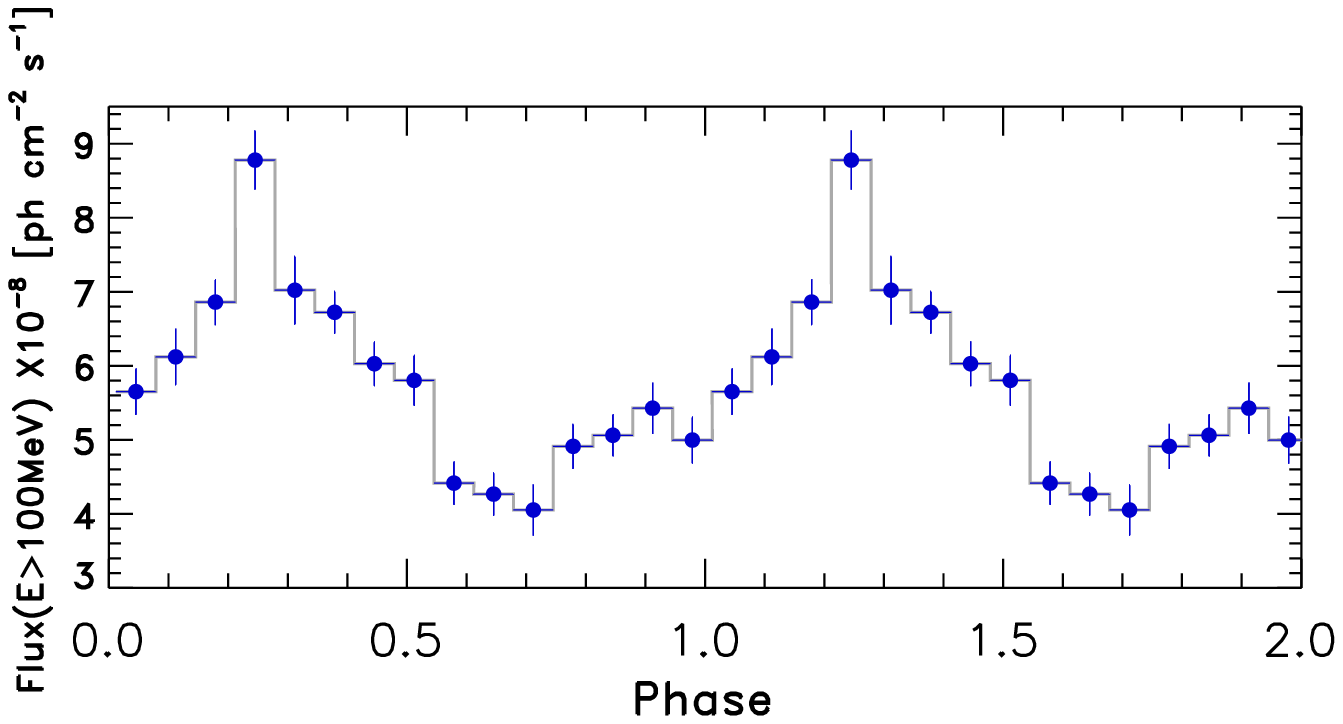}}}\\[-2.7cm]
\hspace{-1.5cm}
\resizebox{1.3\hsize}{!}{\rotatebox[]{90}{\includegraphics{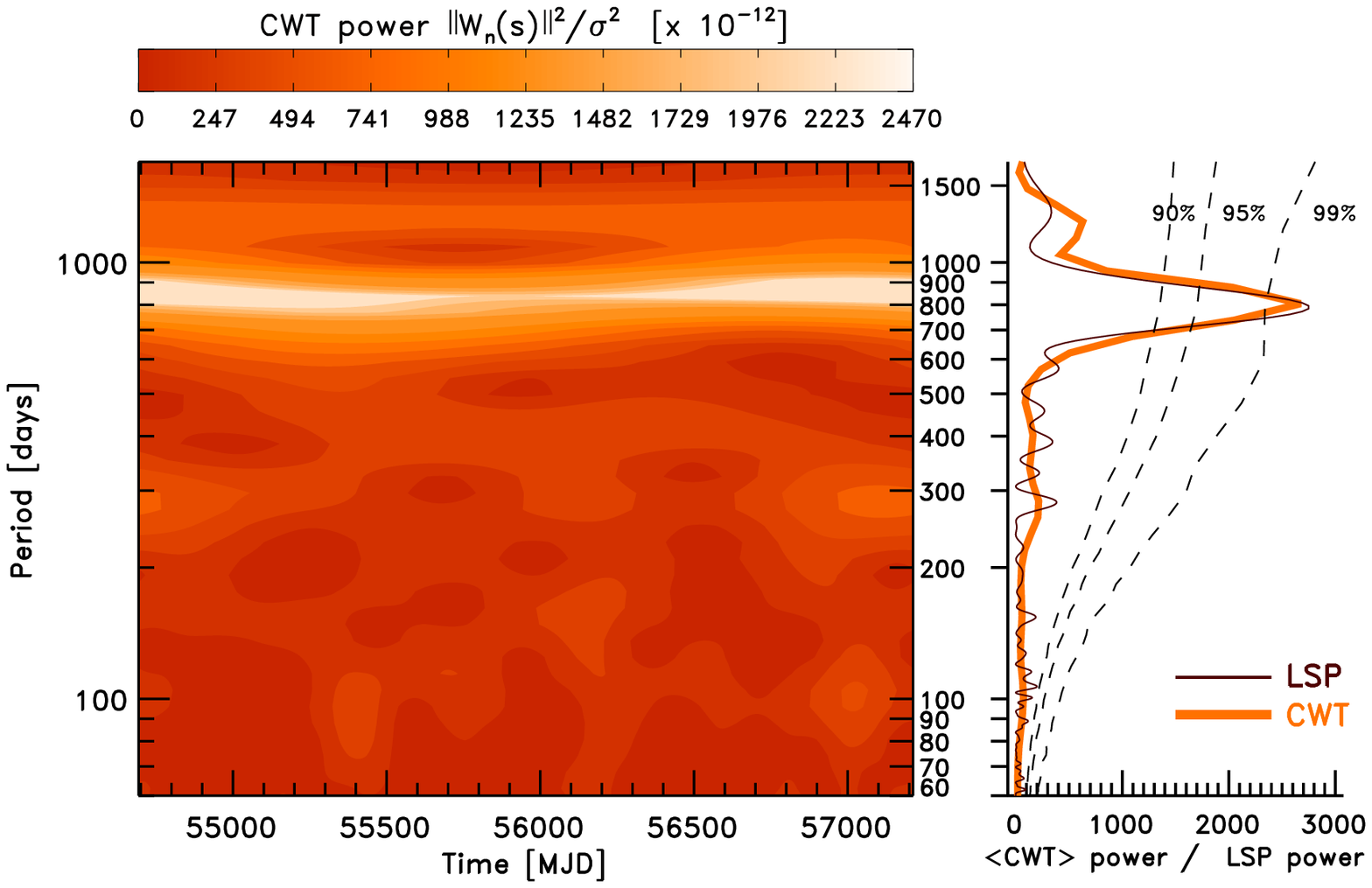}}}
\end{minipage}%
\begin{minipage}[c]{0.5\linewidth}
\hspace{-0.1cm}
\resizebox{1.0\hsize}{!}{\rotatebox[]{0}{\includegraphics{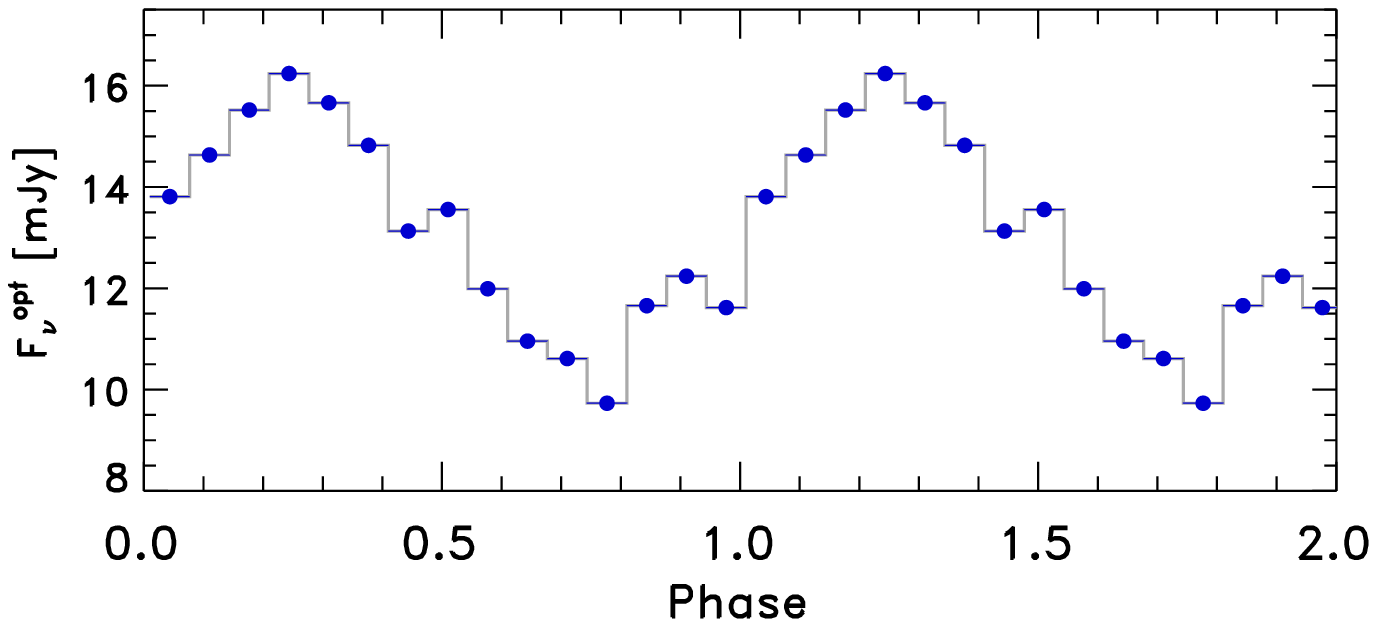}}}\\[-2.7cm]
\hspace{0.6cm}
\resizebox{1.3\hsize}{!}{\rotatebox[]{90}{\includegraphics{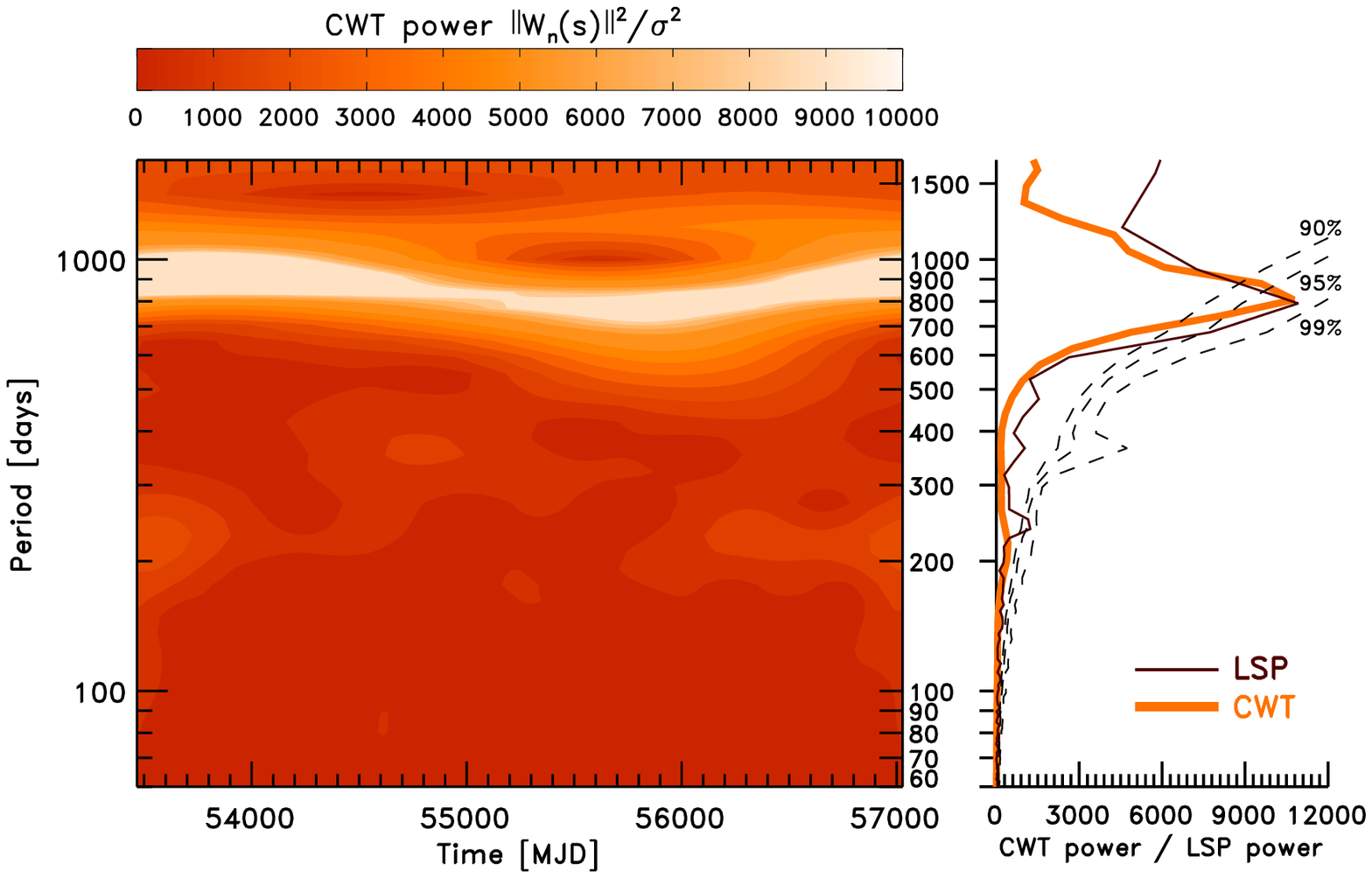}}}
%
\end{minipage}
%
%
%
%
%
\vskip -1.0cm \caption{
\footnotesize{
Left top panel: pulse shape (epoch-folded) $\gamma$-ray ($E>100$ MeV) flux lightcurve at the 2.18 year period (two cycles shown).
Left bottom panels: 2D plane contour plot of the CWT power spectrum (scalogram) of the $\gamma$-ray lightcurve, using a Morlet mother function (filled color contour).
The side panel to this is the 1D smoothed, all-epoch averaged, spectrum of the CWT scalogram showing a signal power peak in agreement with the 2.18-year value, also showing the LSP.
Dashed lines depict increasing levels of confidence against red-noise calculated with Monte Carlo simulation.
The $\gamma$-ray signal peak is above the 99\% confidence contour level  ($<1$\% chance probability of being spurious).
%
Right top panel: pulse shape from epoch folding of the optical flux lightcurve at the 2.18 year period (two cycles shown).
Right bottom panels: the same CWT and LSP diagrams for the optical lightcurve.
The optical signal peak is above the 95\% confidence contour level.
%
}
}
\label{fig:waveletandperiodgram}
\end{figure*}
%

\section{Temporal Variability Analysis and Cross Correlation Analysis}
\label{sect:timing}




We performed continuous wavelet transform (CWT) and Lomb-Scargle Periodogram (LSP) analyses on the lightcurves.
 Fig. \ref{fig:waveletandperiodgram} shows clear peaks at $\sim 2$  years for $\gamma$-ray and optical power spectra.
We also made an epoch folding (pulse shape) analysis used to extract the period, shape, amplitude and phase, with uncertainties \citep{larsson96}.
The $\chi^2$ for the folded pulse as a function of trial periods was fitted with a model containing 4 Fourier components, giving a period of $798 \pm 30$ days ($2.18 \pm 0.08$ years), consistent with the CWT and LSP findings (Fig. \ref{fig:waveletandperiodgram}).
The value of the signal power peak does not change using regular 20-day and 45-day bins or an adaptive-bin technique \citep{lott12} for construction of the LAT lightcurve.

\begin{figure}[hhhht!!]
\centering 
\hskip -0.9cm
\resizebox{1.1\hsize}{!}{\rotatebox[]{-90}{\includegraphics{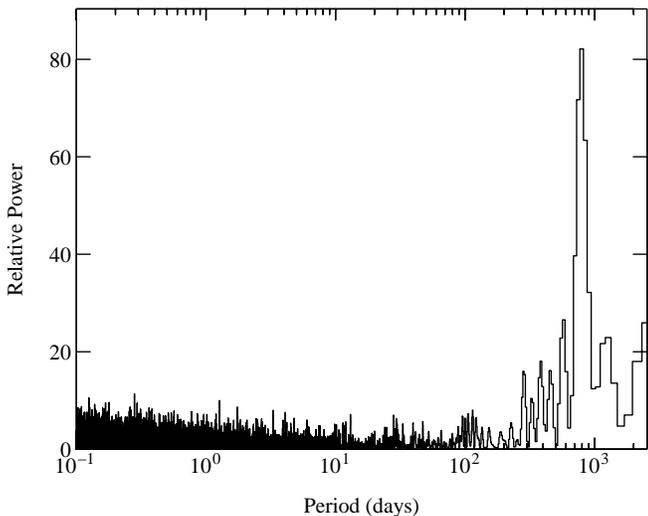}}}
\vskip -1.4cm
\caption{ %
\footnotesize{
Power Density Spectrum (PDS) of the LAT $0.1-300$ GeV count rate
lightcurve of \pg\ from a 3$^{\circ}$ exposure-weighted aperture photometry technique with 600-second time bins.
%
%
%
}
} \label{fig:aplc_psd}
\vskip -0.3cm
\end{figure}

A direct Power Density Spectrum (PDS) constructed from a LAT count-rate lightcurve using exposure-weighted aperture photometry \citep{corbet07,kerr11} above 100 MeV
for a region with 3$^{\circ}$ radius
with 600 second time bins (Fig. \ref{fig:aplc_psd}), confirms previous results with a peak at $2.16 \pm 0.08$ years, at $82\times$ the mean power level.
The low-frequency modulation prevents an easy fit subtraction to the PDS continuum.
The peak is $\sim 5$ times the mean level using a 4th order polynomial fit.


The significance of any apparent periodic variation
depends on what assumption is made about spurious stochastic variability mimicking a periodic variation. The significance of the
$\sim 2$-year $\gamma$-ray periodicity
is difficult to assess given the limited length of the $\gamma$-ray lightcurve.
Red-noise, i.e. random and relatively enhanced low-frequency fluctuations over intervals comparable to the sample length,
hinders the evaluation of periodicity significance \citep[e.g. ][]{hsieh05,lasky15}.
We have approached the problem with two procedures:



1) The red-noise is assumed to be produced by similar amplitude flares (as seen in \pg\ and some other LAT blazars),
and the probability for these to line up in a regular pattern is estimated. The coherence of the periodic modulation was investigated by studying phase variations along the lightcurve. The local phase at each minimum and maximum was estimated by correlating a one-period long data segment with the Fourier template of the full lightcurve. The rms variations relative to a perfectly coherent modulation was 27.4 days.
%
The chance probabilities for 3, 4 and 5 random events to be distributed with at least this coherence, as estimated by Monte Carlo simulations, are 0.0535, 0.0105 and 0.0027 respectively, implying a chance probability of a few percent for the 3.5-peak $\gamma$-ray lightcurve of \pg.
%

%
%
%
%
%
%


2) We modeled the red-noise using Monte Carlo simulations with a first-order autoregressive
process as the null hypothesis to assess whether the signal is consistent with a stochastic origin. Non-linear influence on the PDS is minimal thanks to the evenly spaced $\gamma$-ray lightcurve. The power peak in Fig. \ref{fig:waveletandperiodgram} is above the 99\% confidence contour level, i.e. has $<1\%$ chance of being a statistical fluctuation.
The optical power peak has $<5\%$ chance of being a statistical fluctuation.


%

%
%
%
%

Although the $\gamma$-ray periodicity signal alone is not compelling, the 9.9-years of optical data support the finding of a periodic oscillation in \pg.
%
%
The optical data,
although affected by seasonal gaps, were analyzed using the same techniques as for the $\gamma$-ray data.  This analysis gives a period of $754 \pm 20$ days ($2.06 \pm 0.05$ years), consistent within uncertainties with the $\gamma$-ray results (Fig. \ref{fig:waveletandperiodgram}).
%
%

%

The less coherent 15 GHz lightcurve (5.7-years OVRO data) shows a signal power peak at $1.9 \pm 0.1$ year, with an additional power component at a 1.2-year timescale. 
%
\emph{Swift} XRT data show a factor of 5 variation linearly correlated with the $\gamma$-ray flux, while the synchrotron peak frequency shows a factor $\sim 6$ increase during high X-ray states, as suggested by \citet{reimer08}.
%

The long-term X-ray count rate lightcurve from the \emph{Rossi}-XTE ASM instrument (1996 February 20 to 2010 September 11) and the \emph{Swift}-BAT (from 2005 May 29) were also analyzed but do not show any signal above the low-frequency noise, because of insufficient statistics.

\begin{figure}[hhhhhhtt!!]
\centering 
\vskip -0.1cm
\resizebox{1.0\hsize}{!}{\rotatebox[]{0}{\includegraphics{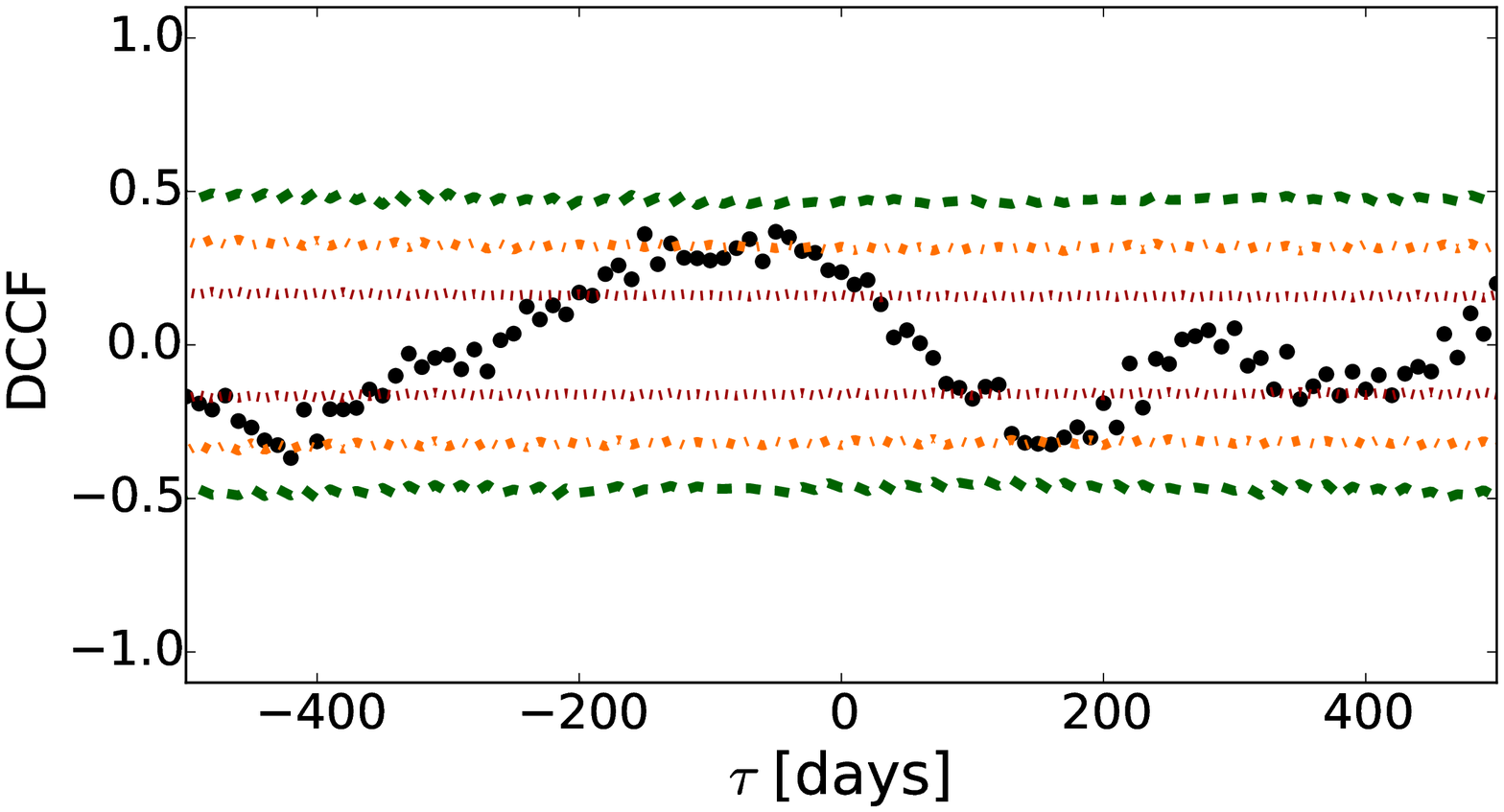}}}\\[0.1cm] 
\vskip -0.3cm
\resizebox{1.0\hsize}{!}{\rotatebox[]{0}{\includegraphics{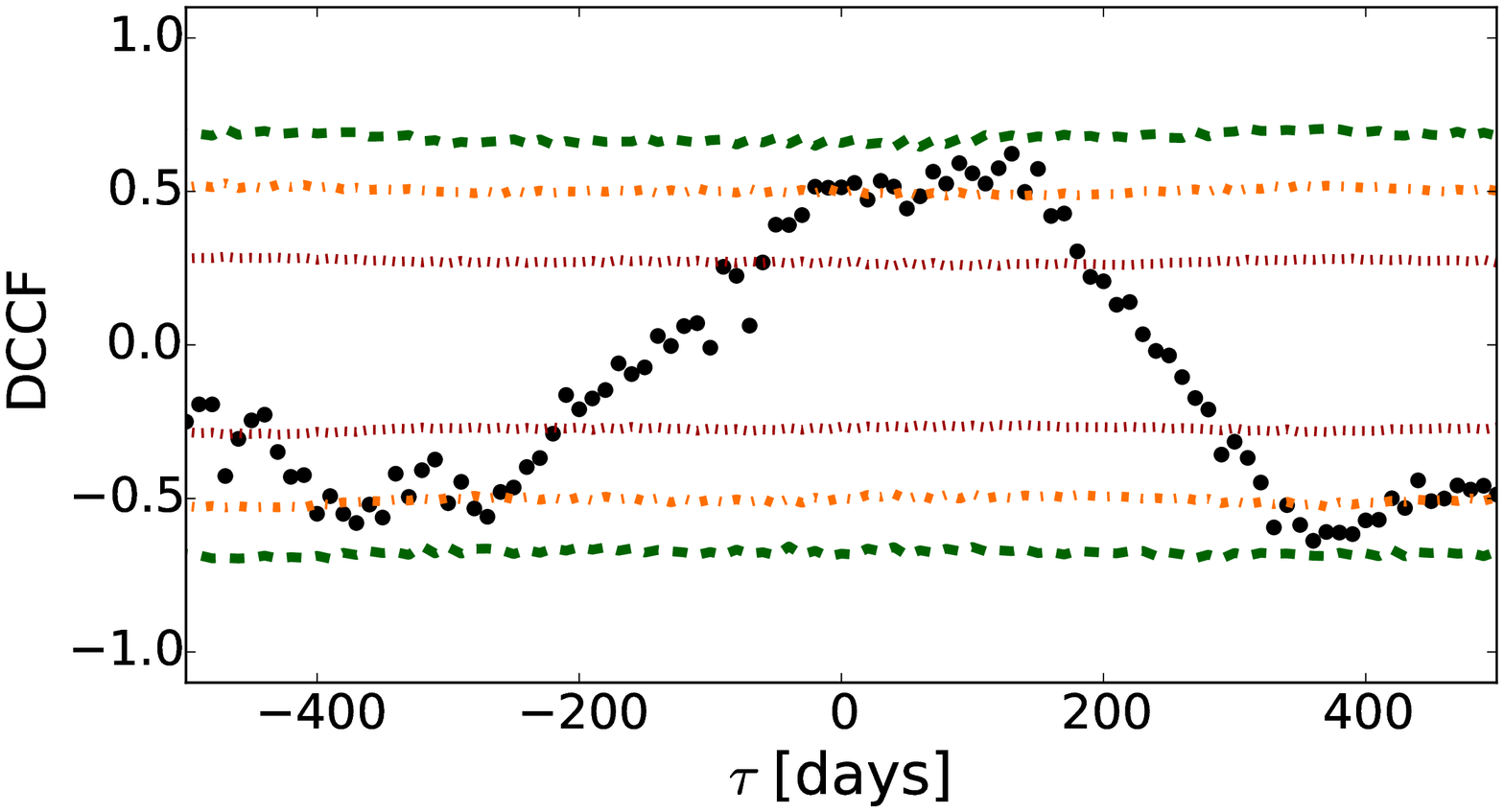}}}\\[0.6cm]  
\vskip -0.6cm
\resizebox{1.0\hsize}{!}{\rotatebox[]{0}{\includegraphics{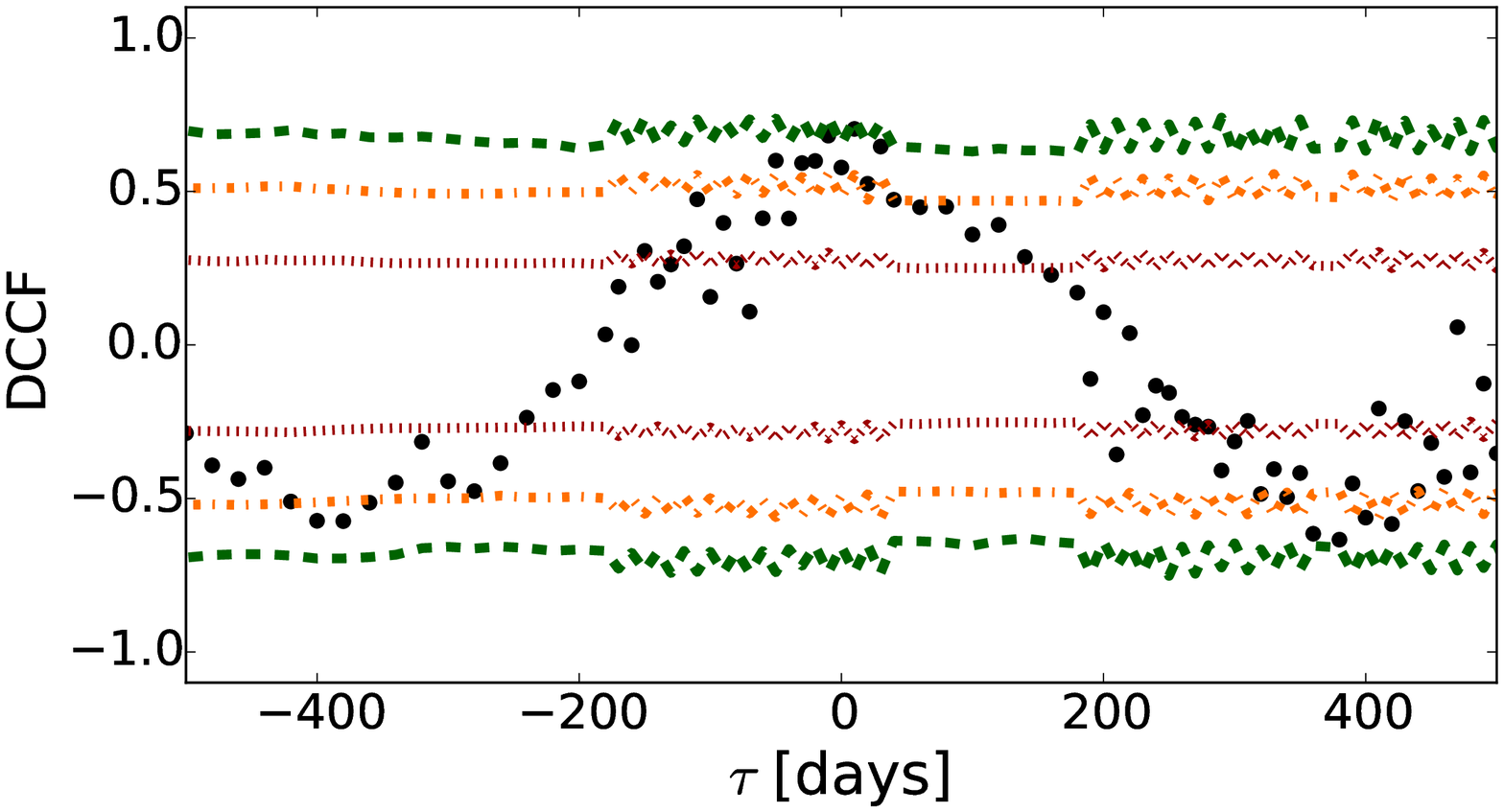}}}
\vskip -0.3cm \caption{%
\footnotesize{Discrete cross-correlation plots from the approach with
PDS model measured from the lightcurve data \citep{maxmoerbeck14}. In each plot the black dots are the DCCF estimates, and the red, orange and green lines are the  1$\sigma$, 2$\sigma$ and 3$\sigma$ significance levels respectively. Top panel: DCCF between the radio 15GHz and $\gamma$-ray (20-day time bins) lightcurve.
Central panel: DCCF between the unbinned optical lightcurve and $\gamma$-ray (20-day time bins) lightcurve.
Bottom panel: DCCF between the 20-day rebinned optical lightcurve and $\gamma$-ray (20-day time bins) lightcurve.
The oscillating shape of the significance contours for this case is due to the number of samples in each bin.
}
}
\label{fig:dccf}
\end{figure}


An important diagnostic
%
%
for multi-frequency periodicity analysis is the discrete cross-correlation function (DCCF) used with two independent and complementary approaches.

In the first procedure, flux variations are modeled assuming a simple power law $\propto 1/f^{\alpha}$ (with $f=1/t$) in the PDS as measured directly from the lightcurve data, allowing us to estimate the cross-correlations significance avoiding the assumption of equal variability in all sources at the cost of a model assumption \citep{maxmoerbeck14}.
For the $\gamma$-ray  lightcurve with 20-day binning we obtain a best fit $\alpha=0.8$, but the error is unconstrained, indicating that the length of the data set is too short (i.e. below five cycles), relative to the suspected periodic modulation, to enable a reliable data characterization.  The 45-day bin lightcurve yields a best fit $\alpha=0.1$ with unconstrained error.
%
%
%
%
%
%
The optical PSD is constrained: the best fit value is $\alpha=1.85$, with 1$\sigma$ limits at $[1.75,2.00]$. The 15 GHz flux light curve a slope of $\alpha = 1.4 $, with unconstrained limits on the $\alpha $ values as for the $\gamma$-ray data.
%
The DCCF between the unbinned radio lightcurve and the 20-day bin $\gamma$-ray lightcurve results in a most probable time lag for radio-flux lagging the $\gamma$-ray flux by $50 \pm 20$ days, with a 98.14\% significance for the best PSD fit  with a range of [89.56\%-99.99\%] when fit errors are taken into account (Fig. \ref{fig:dccf}), using the fitting procedure of \citet{maxmoerbeck14}.  The DCCF between the unbinned optical lightcurve and the 20-day bin $\gamma$-ray lightcurve results in a most probable time lag for $\gamma$-ray flux lagging the optical flux by  $130 \pm 14$ days,
with a 99.14\% significance for the best PSD fit and [96.09\%-99.97\%] when fit errors are taken into account (Fig. \ref{fig:dccf}).  The DCCF peak is broad, however, and consistent with no lag. This is also seen when the optical data are rebinned into 20-day intervals, as shown in the bottom panel, where the most probable lag is $ 10 \pm 51$ days.


In the second procedure,
the significance of the $\gamma$-ray -- radio correlation was estimated to be 95\% by a mixed source correlation procedure \citep{fuhrmann14}, cross-correlating the \pg~lightcurve with those of 132 comparison sources in that work, and evaluating the average DCCF level for time lags $-100$ to $+100$ days. The $\gamma$-ray -- optical correlation is significant at the 99\% level, even though partly limited by the number of comparison sources and optical lightcurve gaps.
%
%
With only 132 comparison lightcurves we can measure a minimum probability-value of 0.0075, therefore in principle a 99\% level of significance, but in this approach the error in that estimate is hard to determine.
With the mixed source methods there are two limitations: 1) the assumption that all the sources can be described with the same model for the variability, and 2) the sample variance due to the limited number of lightcurves  must be assessed.
The optical flux is found to lead the $\gamma$-ray variations by $75 \pm 27$ days and the radio by $158 \pm 10$ days ($\gamma$-ray variations lead the 15 GHz flux variations by $83 \pm 27$ days). The possible reverse $\gamma$-ray-optical time lag decreases to $28\pm 27$ days
when the optical lightcurve is binned. 
%
%

The possible optical-$\gamma$-ray lag was already pointed out by \citet{cohen14}, using KAIT unbinned optical lightcurves and LAT data.
%
%
%
The high degree of $\gamma$-ray-radio correlation in \pg\ is not typically found in other individual blazars/AGN \citep[see][]{maxmoerbeck14}. Significant cross-correlations are, nevertheless, found when stacking blazar samples \citep[radio lagging $\gamma$ rays; ][]{fuhrmann14}.

%
%

\section{Discussion and Conclusions} \label{sect:discussion}

Factors that led to the indication of a possible $\sim 2$-year periodic modulation in \pg\ are: the continuous all-sky survey of \textit{Fermi};
the increased capability of the new \fermi LAT Pass 8 data; and the
%
long-term radio/optical monitoring of $\gamma$-ray blazars.  Although the statistical significance of periodicity is marginal in each band, the consistent positive cross-correlation between bands strengthens the case, making \pg\ the first possible quasi-periodic
GeV $\gamma$-ray blazar and a prime candidate for further studies. Hints of possible $\gamma$-ray periodicities are rare in literature
\citep[for example ][]{sandrinelli14}.
The similarity of the low- and high-energy  modulation in \pg\ is also a novel behavior for AGN \citep[][]{rieger04,rieger07}.
%
%
%
%
Any periodic driving scenario should be related to the relativistic jet itself or to the process feeding the jet for this VHE BL Lac object. We outline, as examples, four possibilities:
%
%

\begin{enumerate}

\item
Pulsational accretion flow instabilities, approximating periodic behavior, are able to explain modulations in the energy outflow efficiency. Magnetically-arrested and magnetically-dominated accretion flows (MDAF) could be suitable regimes for radiatively inefficient BL Lacs \citep{fragile09}, characterized by advection-dominated accretion flows and subluminal, turbulent and peculiar radio kinematics \citep{karouzos12,piner14}. Such kinematics are sometimes explained as a precessing or helical jet \citep{conway93}. MDAF in a inner disc portion can be able to efficiently impart energy to particles in the jets of VHE BL Lacs \citep{tchekhovskoy11}.
Periodic instabilities are believed to have short periods, $\sim 10^5\,$s $\cdot$ ($M_{SMBH}$/$10^8\,$M$_{\odot})$ \citep{honma1992}, but MHD simulations of magnetically choked accretion flows are seen to produce longer periods for slow-spinning SMBH \citep{mckinney12}.
%

\item
Jet precession \citep[e.g., ][]{romero00,stirling03,caproni13}, rotation \citep{camezind92,vlahakis98,hardee99} or helical structure \citep[e.g., ][]{conway93,roland94,villata99,nakamura04,ostorero04}, i.e. geometrical models \citep{rieger04}, in the presence of a jet wrapped by a sufficiently strong magnetic field, could have a net apparent periodicity from the change of the viewing angle.
Correspondingly the resulting Doppler magnification factor changes periodically without the need for intrinsic variation in outflows and efficiency.
%
Non-ballistic hydrodynamical jet precession may explain variations with periods $> 1$ year \citep{rieger04}.
A differential Doppler factor $\Delta \mathcal{D}(t) = \Gamma^{-1}(1- \beta(t) \cos \theta(t))^{-1} \lesssim 40\%$ variation (precession angle $\sim 1\arcdeg$) might be sufficient to support the  $\sim 2.8$ amplitude flux modulation seen in $\gamma$ rays.
A homogeneous curved helical jet scenario for \pg\ was proposed in \citet{raiteri15}.
\item

A mechanism analogous to low-frequency QPO from Galactic high-mass binaries/microquasars could produce
an accretion-outflow coupling mechanism as the basis of the periodicity \citep{fender04}.
 \citet{king13} ascribed the radio QPO in the FSRQ CGRaBS J1359+4011 to this mechanism. However BL Lac objects like \pg\ are thought to possess a lower accretion rate. The microquasar QPO mechanism of Lense-Thirring precession \citep{wilkins72} requires that the inner accretion flow forms a geometrically thick torus rather than a standard thin disc as the latter warps \citep[Bardeen-Petterson effect, ][]{bardeen75} rather than precesses
\citep{ingram09}.
A low mass accretion rate means that the
accretion process probably forms an Advection-Dominated Accretion Flow (ADAF), so it can precess \citep{fragile09}. The X-ray emission in \pg\ is probably from the jet rather than from the flow, making
it unlikely that the changing inclination of the hot flow causes the QPO. However,
Lense-Thirring precession of the flow could affect the jet direction,
giving the QPO as in (2) above.

\item
The presence of a gravitationally bound binary SMBH system \citep{begelman80,barnes92} with a total mass $\sim 10^8$ M$_{\odot}$, and a milli-pc separation in the early inspiral gravitational-wave driven regime, might be another hypothesis.
Keplerian binary orbital motion, would induce periodic accretion perturbations \citep{valtonen08,pihajoki13,liu15}
or jet nutation expected from the misalignment of the rotating SMBH spins, or the gravitational torque on the disc exerted by the companion \citep{katz97,romero00,caproni13,graham15}.
Significant acceleration of the disc evolution and accretion onto a binary SMBH system is depicted by modeling \citep{nixon13,dogan15}
.

%
%
%
%
%
%
Binary SMBH induced periodicities have timescales ranging from $\sim 1$ to $\sim 25$ years  \citep{komossa06,rieger07}.
The SMBH total mass in \pg, estimated utilizing the putative link between inflow/accretion (disc luminosity) and outflow/jet (jet power) in blazars \citep{ghisellini14}, is $\simeq 1.6\times 10^8$ M$_{\sun}$, using a 0.1 $\dot M_{Edd}$ rate and Doppler factor $\mathcal{D}=30$,
%
%
in agreement with estimates for VHE BL Lacs \citep{woo05}.

%
%
The observed 2.18-year period is
equivalent to an intrinsic orbital time $T'_{Kep} \leq T_{obs}/(1+z) \simeq 1.5 $ years, and
%
%
the binary system size would be  $0.005$ pc ($ \sim 100$ Schwarzschild radii).
%
%
The probability to be observing such milli-pc system, estimated from the binary mass ratios $\sim 0.1 - 0.01$ and the GW-driven regime lifetime \citep{peters64}, $t_{GW}
\simeq 10^5 - 10^6 $ years
might be too small.
%
%
%
%
%
%

\end{enumerate}

%
%
Periodicities claimed for AGN are often controversial;
%
%
%
however \pg\ may potentially represent a key $\gamma$-ray/multimessenger laboratory in the hypothesis of low-frequency gravitational wave emission and may have associated PeV neutrino emission \citep{padovani14}.
%
%
%
VLBI structure observations, radio/optical polarization data, and a prolonged multifrequency monitoring campaign will shed light on the situation.
If the periodic modulation is real and coherent, as would be expected for a binary scenario, then subsequent maxima would be expected in 2017 and 2019, well within the possible lifetime of the \fermi mission.



\acknowledgments

\footnotesize{
We thank the anonymous referee for useful and constructive comments. We extend special thanks to Prof. C. Done of Durham University, UK, and Prof. R. W. Romani of Stanford University, USA, for useful comments during the course of this work.
The \textit{Fermi}-LAT Collaboration acknowledges support for LAT development, operation and data analysis from NASA and DOE (United States), CEA/Irfu and IN2P3/CNRS (France), ASI and INFN (Italy), MEXT, KEK, and JAXA (Japan), and the K.A.~Wallenberg Foundation, the Swedish Research Council and the National Space Board (Sweden). Science analysis support in the operations phase from INAF (Italy) and CNES (France) is also gratefully acknowledged.
Tuorla blazar monitoring program has been partially supported by Academy of Finland grant 127740. KAIT telescope program is supported by Katzman Foundation and the National Science Foundation. The CSS survey is funded by the National Aeronautics and Space Administration under Grant No. NNG05GF22G issued through the Science Mission Directorate Near-Earth Objects Observations Program. The CRTS survey is supported by the U.S.~National Science Foundation under grants AST-0909182. The OVRO 40-m program is supported in part by NASA grants NNX08AW31G and NNX11A043G and NSF grants AST-0808050 and AST-1109911. The MOJAVE program is supported under NASA-Fermi grant NNX12A087G.
The National Radio Astronomy Observatory (NRAO) is a facility of the National Science Foundation operated under cooperative agreement by Associated Universities, Inc.
The NASA \textit{Swift} $\gamma$-ray burst explorer is a MIDEX Gamma Ray Burst mission led by NASA with participation of Italy and the UK. This research has made use of  the Smithsonian/NASA's ADS bibliographic database.  This research has made use of the NASA/IPAC NED database (JPL CalTech and NASA, USA). This research has made use of the archives and services of the ASI Science Data Center (ASDC), a facility of the Italian Space Agency (ASI Headquarter, Rome, Italy). This research has made use of the XRT Data Analysis Software (XRTDAS) developed under the responsibility of the ASDC. This work is a product of the ASDC Fermi team developed in the frame of the INAF Senior Scientists project and the foreign visiting scientists program of ASDC.

Facilities: \textit{Fermi Gamma-ray Space Telescope}  --- \textit{Swift}  --- \textit{OVRO}  --- \textit{Tuorla} --- \textit{KVA} --- \textit{KAIT} --- \textit{CSS} --- \textit{CRTS}
}

\normalsize

\bibliographystyle{apj}

{}

\end{document}